# Impact of Nuclear Reaction Uncertainties on AGB Nucleosynthesis Models

S. Bisterzo*,[ab] R. Gallino,[b] F. Käppeler,[c] M. Wiescher,[d] and C. Travaglio[a]

[a]*INAF - Astronomical Observatory Turin, Turin, Italy*
[b]*University of Turin, Physics Department, Turin, Italy*
[c]*Karlsruhe Institute of Technology (KIT), Institut fuer Kernphysik, Karlsruhe, Germany*
[d]*Joint Institute for Nuclear Astrophysics (JINA), Physics Department, University of Notre Dame, Notre Dame, IN*
*E-mail:* bisterzo@to.infn.it, sarabisterzo@gmail.com

Asymptotic giant branch (AGB) stars with low initial mass ($1 < M/M_\odot < 3$) are responsible for the production of neutron-capture elements through the main s-process (main slow neutron capture process). The major neutron source is $^{13}$C($\alpha$, n)$^{16}$O, which burns radiatively during the interpulse periods at $\sim 8$ keV and produces a rather low neutron density ($10^7$ n/cm$^3$). The second neutron source $^{22}$Ne($\alpha$, n)$^{25}$Mg, partially activated during the convective thermal pulses when the energy reaches about 23 keV, gives rise to a small neutron exposure but a peaked neutron density ($N_n$(peak) $> 10^{11}$ n/cm$^3$). At metallicities close to solar, it does not substantially change the final s-process abundances, but mainly affects the isotopic ratios near s-path branchings sensitive to the neutron density.

We examine the effect of the present uncertainties of the two neutron sources operating in AGB stars, as well as the competition with the $^{22}$Ne($\alpha$, $\gamma$)$^{26}$Mg reaction. The analysis is carried out on the main-s process component (reproduced by an average between $M_{\rm ini}^{\rm AGB}$ = 1.5 and 3 $M_\odot$ at half solar metallicity, see [3]), using a set of updated nucleosynthesis models. Major effects are seen close to the branching points. In particular, $^{13}$C($\alpha$, n)$^{16}$O mainly affects $^{86}$Kr and $^{87}$Rb owing to the branching at $^{85}$Kr, while small variations are shown for heavy isotopes by decreasing or increasing our adopted rate by a factor of 2–3. By changing our $^{22}$Ne($\alpha$, n)$^{25}$Mg rate within a factor of 2, a plausible reproduction of solar s-only isotopes is still obtained. We provide a general overview of the major consequences of these variations on the s-path. A complete description of each branching will be presented in Bisterzo et al., in preparation.

*Speaker.

## 1. Introduction

Asymptotic giant branch stars (AGBs) manufacture about half of the heavy elements from Sr to Pb-Bi (see e.g., [5, 18, 39]) through the main component of the s process (slow neutron capture process), when they climb for the second time the red giant branch and experience a series of He-shell flashes called thermal pulses (TPs). The s-process abundances observed today in the Solar System are the result of a complex Galactic evolution mechanism, which accounts for the pollution of several AGB generations with different initial masses and metallicities. Galactic chemical evolution models are needed to properly interpret the dynamics of the s-process in the Milky Way ([44, 46]). However, it was shown that, as a first approximation, AGB stars with low initial masses and half solar metallicity may reproduce the main s-process component ([3]). The simulations were made by considering an average between AGB stellar models with initial masses of $M$ = 1.5 and 3 $M_\odot$ at [Fe/H] = $-0.3$, and a specific $^{13}$C-pocket choice (called 'case ST'). This approximation is useful to test the impact of the present nuclear cross section uncertainties on the main-s component with an updated network. In Fig. 1, we show the solar main s-process component computed with an updated network as described by [4, 22], further improved with new cross section measurements of $^{92,94,96}$Zr [40, 41, 42], $^{186,187,188}$Os [32], $^{64,70}$Zn [36], Mg isotopes [31]). A plausible reproduction of all s-only isotopes (full circles) is obtained within the uncertainties.

We examine the effect of the uncertainties of the two AGB neutron sources, $^{13}$C($\alpha$, n)$^{16}$O and $^{22}$Ne($\alpha$, n)$^{25}$Mg (as well as of the $^{22}$Ne($\alpha$, $\gamma$)$^{26}$Mg reaction) on the main-s process component. Both reaction rates are influenced by the unknown contribution of subthreshold states (e.g., the state at 6.356 MeV for $^{13}$C($\alpha$, n)$^{16}$O) and resonances (the 635 keV resonance for $^{22}$Ne($\alpha$, n)$^{25}$Mg). Several experimental efforts have been made in the past years to reduce the uncertainties of these two reactions, e.g., see [20, 27, 25, 33, 17, 15][1] for $^{13}$C($\alpha$, n)$^{16}$O, and [21, 2, 19, 26, 23, 47, 29] for $^{22}$Ne($\alpha$, n)$^{25}$Mg. In Sections 2, 3, and 4, we describe the test carried out starting from our adopted rates, and we present the major effects on heavy isotopes with atomic masses from 70 to 210.

## 2. $^{13}$C($\alpha$, n)$^{16}$O

For the $^{13}$C($\alpha$, n)$^{16}$O reaction, we adopt the rate by [12], which is close to the recent values provided by [17, 15]. Estimations by [2, 27] suggest that, at $\sim$8 keV, the $^{13}$C($\alpha$, n)$^{16}$O rate can vary within a factor of 2 or 3 with respect to our rate. Starting from these uncertainties, we carried out two tests on the main-s component: test **[$^{13}$C($\alpha$, n)$^{16}$O]/3** (lower limit by [27]), test **[$^{13}$C($\alpha$, n)$^{16}$O]$\times$2** (upper limit by [2]).
In our AGB models (based on the prescriptions provided by FRANEC stellar models [37, 38]) all $^{13}$C burns radiatively in the pocket during the interpulse[2]. Therefore, an amount of $^{13}$C (in mass fraction X($^{13}$C) $\lesssim 10^{-5}$) that is negligible for the s-process is ingested in the next thermal pulse, even by adopting the lower limit by [27].

---

[1] Previous measurements show even larger uncertainties ([6, 13, 12, 16, 2].

[2] Note that new FRANEC stellar models [8, 10] experience a partial convective burning of $^{13}$C during the first (or second) formed $^{13}$C-pocket. This occurs in metal-rich models and may influence some isotopic ratios close to the branchings (see also [7]). However, the effect on the final s distribution should be small because of the contribution of the following standard $^{13}$C-pockets. On the other hand, at low metallicities some protons can be ingested in the He-intershell during the first fully developed Thermal Pulse, leading to a significative s-process nucleosynthesis (see [9] and references therein). Both phenomena have been recently confirmed and analysed in detail by [30]. In particular, [30] found that the partial convective burning of $^{13}$C is more important at low metallicity.

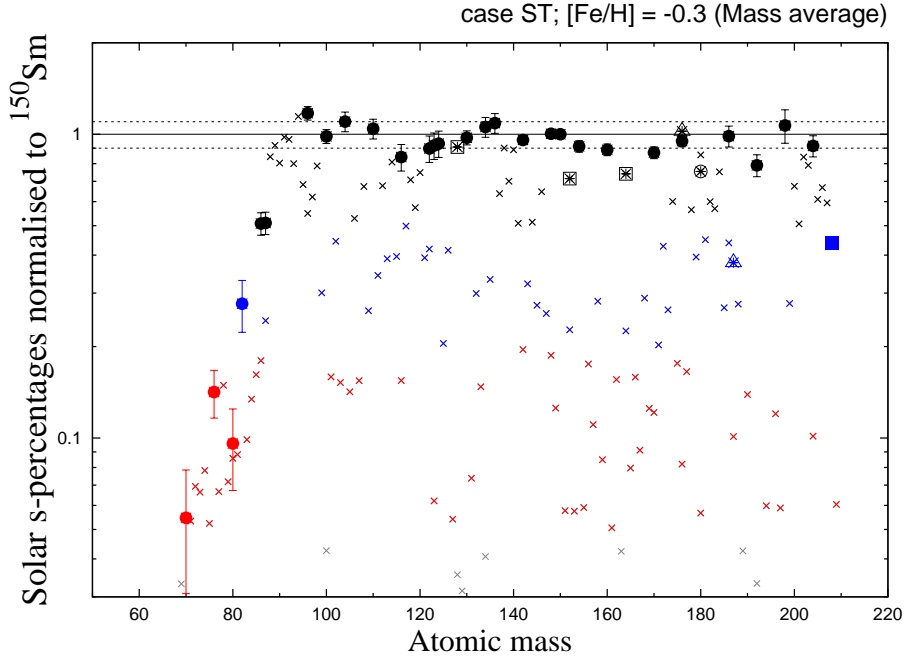

**Figure 1:** The solar main s-process component versus atomic mass is reproduced by assuming a ST $^{13}$C-pocket, half solar metallicity, and by averaging between AGB models of initial masses $M$ = 1.5 and 3 $M_\odot$ (as in [3], but updated to 2012, see text). The s-production factors in the He-intershell (given in mass fraction '$X_i$' over the solar-scaled initial values) normalised to $^{150}$Sm are represented. As in [22], different symbols are used for isotopes that receive additional non-negligible contributions: $^{128}$Xe, $^{152}$Gd, and $^{164}$Er (p contribution; crossed squares); $^{176}$Lu and $^{187}$Os (crossed triangles) the first is a long-lived isotope that decays into $^{176}$Hf, while the second is affected by the long-lived decay of $^{187}$Re; $^{180}$Ta (crossed circle), which also receives contributions from the p process and from $\nu$-nucleus interactions in massive stars; $^{208}$Pb (filled square), which should receive an additional half contribution by the strong s-process component [45]. We represent with black symbols isotopes that are mainly produced by the s-process ($\geq$50%), blue symbols isotopes produced for 20–50% by the s-process, red symbols for 5–20% and grey symbols for negligible s-contribution (<5%). This will be useful for Section 3. The error bars displayed for the s-only isotopes account for the uncertainties of the solar meteoritic abundances by [1], with the exception of Xe isotopes, for which we adopted [28].

The main-s process component is marginally affected by both tests. The variations among isotopes heavier than $A$ = 90 are $\lesssim$1%. The uncertainty of the $^{13}$C($\alpha$, n)$^{16}$O reaction mainly affects isotopes close to the branching at $^{85}$Kr, which is sensitive to the neutron density. In particular, $^{86}$Kr and $^{87}$Rb decrease by 27% and 14% with the test [$^{13}$C($\alpha$, n)$^{16}$O]/3, and increase by 15% and 7% with the test [$^{13}$C($\alpha$, n)$^{16}$O]×2, respectively. Note that both isotopes receive a low contribution from the main-s process component, $\lesssim$20% for $^{86}$Kr and $\lesssim$30% for $^{87}$Rb. The s-only isotopes $^{86,87}$Sr are not affected by the $^{85}$Kr branching (differences $\lesssim$1%). Note that the variation of $^{86}$Kr and $^{87}$Rb increases by decreasing the AGB initial mass, because the additional contribution of the $^{22}$Ne($\alpha$, n)$^{25}$Mg reaction is reduced due to the lower temperatures reached during the TPs.

## 3. $^{22}$Ne($\alpha$, n)$^{25}$Mg

For the $^{22}$Ne($\alpha$, n)$^{25}$Mg reaction we adopt the lower limit by [21], in which the resonance contribution at 633 keV has been neglected. This value is close to that provided by NACRE and to the upper limit by [29]. The large upper limit suggested in 1999 by [2] (up to a factor of 50 at $\sim$23 keV) has been reduced by at least one order of magnitude by the most recent measurements. We carried out two tests, based on the recent uncertainties: test **[$^{22}$Ne($\alpha$, n)$^{25}$Mg]/2** (close to the lower limit by [19, 29]) and test **[$^{22}$Ne($\alpha$, n)$^{25}$Mg]$\times$2** [21]. Two additional extreme cases are also included in the analysis: tests **[$^{22}$Ne($\alpha$, n)$^{25}$Mg]$\times$4** (close to the upper limit by [21]) and **[$^{22}$Ne($\alpha$, n)$^{25}$Mg]/4**. The results of these four tests are represented in Fig. 2.

It is known that the uncertainty of the $^{22}$Ne($\alpha$, n)$^{25}$Mg mainly influences several branchings (e.g., [3]). Among the s-only isotopes major differences are shown by $^{80}$Kr ($<$10% of which is produced by the main-s process), $^{86,87}$Sr and $^{96}$Mo, owing to the branchings at $^{79}$Se, $^{85}$Kr and $^{95}$Zr, respectively. Among the other isotopes important variations are shown by $^{86}$Kr and $^{87}$Rb, $^{96}$Zr (which decreases by $\sim$60% with the test [$^{22}$Ne($\alpha$, n)$^{25}$Mg]/2). The branchings at $^{134,135}$Cs modify the abundances of the two s-only isotopes $^{134,136}$Ba (e.g., 3-4% variations with the test [$^{22}$Ne($\alpha$, n)$^{25}$Mg]/2); in this atomic mass region, $^{135}$Ba shows the largest differences. The branchings at $^{151}$Sm and $^{154}$Eu influence the production of the s-only $^{152,154}$Gd (e.g., 7% variations with the test [$^{22}$Ne($\alpha$, n)$^{25}$Mg]/2). The branching at $^{176}$Lu modifies the $^{176}$Lu/$^{176}$Hf ratio ($\sim$15% with the test [$^{22}$Ne($\alpha$, n)$^{25}$Mg]/2). Additional notable branchings are at $^{115}$Cd, $^{121}$Sn, $^{170}$Tm, $^{185}$W, $^{203}$Hg, $^{204}$Tl. In these atomic mass regions, the isotopes affected by the $^{22}$Ne($\alpha$, n)$^{25}$Mg rate are $^{116}$Cd, $^{122}$Sn, $^{170}$Yb, $^{186}$W and $^{187}$Re, $^{204}$Hg, $^{205}$Tl (up to $^{209}$Bi). Finally, we list some additional branchings close to the neutron rich isotopes: e.g., $^{141}$Ce, $^{147}$Nd, $^{169}$Er, $^{175}$Yb, that influence $^{142}$Ce, $^{148}$Nd, $^{170}$Er, $^{176}$Yb. A complete description of each branching will be presented in Bisterzo et al., in preparation.

In intermediate mass AGBs (IMS) ($4 < M/M_\odot < 8$), the $^{22}$Ne($\alpha$, n)$^{25}$Mg reaction becomes the most efficient neutron source owing to the higher temperature reached during the thermal pulses. Therefore the effect of the $^{22}$Ne($\alpha$, n)$^{25}$Mg uncertainty becomes of fundamental importance, see e.g., [29, 48, 24, 23].
In future studies we will investigate the impact of the neutron source uncertainties on AGB stars with different initial mass and metallicity, including AGBs with intermediate mass.

## 4. $^{22}$Ne($\alpha$, $\gamma$)$^{26}$Mg

As for $^{22}$Ne($\alpha$, $\gamma$)$^{26}$Mg, we adopt the lower limit by [21]. We carried out three tests: test **[$^{22}$Ne($\alpha$, $\gamma$)$^{26}$Mg]$\times$8**, upper estimation by [29] that includes the uncertainty of the two resonances near $E_r^{lab} = 830$ keV ([11]); test **[$^{22}$Ne($\alpha$, $\gamma$)$^{26}$Mg]$\times$2**, upper limit by [21] (close to the upper limit by [29]); test **[$^{22}$Ne($\alpha$, $\gamma$)$^{26}$Mg]/4**, additional extreme test.

Appreciable variations are seen for $^{26}$Mg, which is directly involved in the reaction and has a small neutron capture cross section: it increases by a factor of 3.5 and by 34% for the first two tests, and decreases by 26% for the last test. Therefore, the competition with the $^{22}$Ne($\alpha$, n)$^{25}$Mg neutron source is marginal in this AGB mass range. Minor effects are shown by heavy isotopes, even by considering the upper limit by NACRE (test [$^{22}$Ne($\alpha$, $\gamma$)$^{26}$Mg]$\times \sim$25).

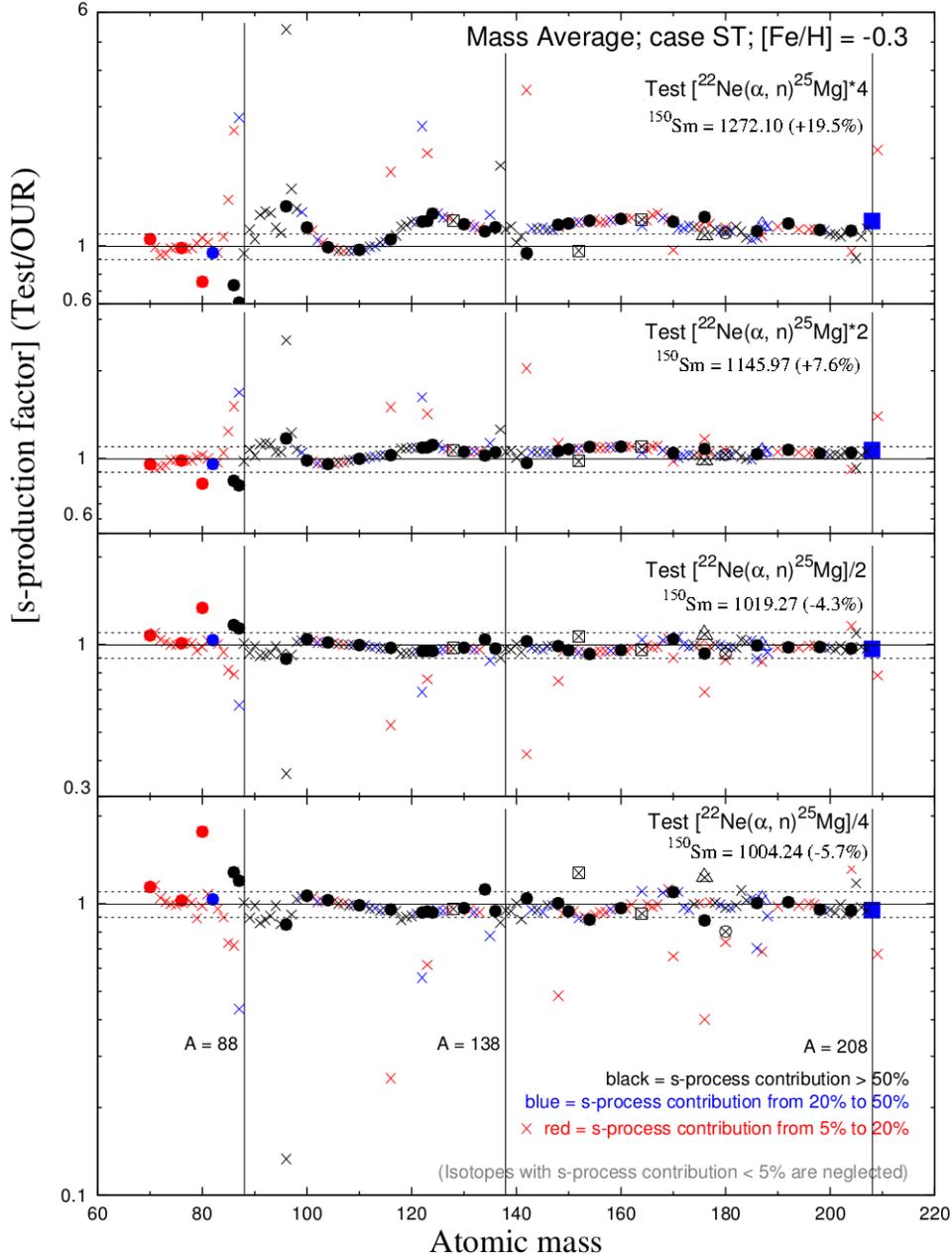

**Figure 2:** We show the ratios between the s-production factors obtained with four $^{22}$Ne($\alpha$, n)$^{25}$Mg tests and our adopted rate ('OUR', shown in Fig. 1): panels a, b, c, d correspond to tests [$^{22}$Ne($\alpha$, n)$^{25}$Mg]$\times$4, $\times$2, /2 and /4, respectively. We considered an average between AGB models of initial masses $M$ = 1.5 and 3 $M_\odot$, a ST $^{13}$C-pocket and half solar metallicity. Symbols are the same as Fig. 1. The variation of $^{150}$Sm with respect to our adopted $^{22}$Ne($\alpha$, n)$^{25}$Mg rate is given in the top-right inset of each panel. The value predicted by the main-s component is $X(^{150}\text{Sm})/X(^{150}\text{Sm})_{\text{ini}}$ = 1064.97. We excluded isotopes with marginal solar s-process contribution (<5%; grey symbols in Fig 1). Note that for each panel different ordinate ranges are shown.

**Acknowledgments.** This work has been supported by the Joint Institute for Nuclear Astrophysics (JINA, University of Notre Dame, USA) and Karlsruhe Institute of Technology (KIT, Germany).